\documentclass[aps,prl,twocolumn,superscriptaddress,showpacs,longbibliography,floatfix]{revtex4-1}

\usepackage{graphicx}
\usepackage{upgreek}

\begin{document}

\title{Vogel-Fulcher-Tammann Freezing of a Thermally Fluctuating Artificial Spin Ice Probed by X-ray Photon Correlation Spectroscopy}


\author{S.~A.~Morley}
\email[Email:~]{S.A.Morley@leeds.ac.uk}
\affiliation{School of Physics and Astronomy, University of Leeds, Leeds LS2 9JT, United Kingdom}

\author{D.~Alba~Venero}
\affiliation{ISIS, STFC Rutherford Appleton Laboratory, Chilton, Didcot OX11 0QX, United Kingdom}

\author{J.~M.~Porro}
\affiliation{ISIS, STFC Rutherford Appleton Laboratory, Chilton, Didcot OX11 0QX, United Kingdom}

\author{S.~T.~Riley}
\affiliation{School of Electronic and Electrical Engineering, University of Leeds, Leeds LS2 9JT, United Kingdom}

\author{A.~Stein}
\affiliation{Center for Functional Nanomaterials, Brookhaven National Laboratory, Upton, New York 11973, USA}

\author{P.~Steadman}
\affiliation{Diamond Light Source, Chilton, Didcot OX11 0DE, United Kingdom}

\author{R.~L.~Stamps}
\affiliation{SUPA, School of Physics and Astronomy, University of Glasgow, Glasgow G12 8QQ, United Kingdom}

\author{S.~Langridge}
\affiliation{ISIS, STFC Rutherford Appleton Laboratory, Chilton, Didcot OX11 0QX, United Kingdom}

\author{C.~H.~Marrows}
\email[Email:~]{C.H.Marrows@leeds.ac.uk}
\affiliation{School of Physics and Astronomy, University of Leeds, Leeds LS2 9JT, United Kingdom}

\date{\today}

\begin{abstract}
  We report on the crossover from the thermal to athermal regime of an artificial spin ice formed from a square array of magnetic islands whose lateral size, 30~nm~$\times$~70~nm, is small enough that they are superparamagnetic at room temperature. We used resonant magnetic soft x-ray photon correlation spectroscopy (XPCS) as a method to observe the time-time correlations of the fluctuating magnetic configurations of spin ice during cooling, which are found to slow abruptly as a freezing temperature $T_0 = 178 \pm 5$~K is approached. This slowing is well-described by a Vogel-Fulcher-Tammann law, implying that the frozen state is glassy, with the freezing temperature being commensurate with the strength of magnetostatic interaction energies in the array. The activation temperature, $T_\mathrm{A} = 40 \pm 10$~K, is much less than that expected from a Stoner-Wohlfarth coherent rotation model. Zero-field-cooled/field-cooled magnetometry reveals a freeing up of fluctuations of states within islands above this temperature, caused by variation in the local anisotropy axes at the oxidised edges. This Vogel-Fulcher-Tammann behavior implies that the system enters a glassy state on freezing, which is unexpected for a system with a well-defined ground state.
\end{abstract}


\maketitle


In the past decade a new species of magnetic metamaterials has emerged: the artificial spin ices (ASI) \cite{Nisoli2013,HandSreview}. They consist of a 2-dimensional array of nanoscale magnetic islands arranged so that the magnetostatic interactions between the islands are geometrically frustrated \cite{wang}. The size and shape of the individual islands are designed with the intention that their shape anisotropy means they act as single-domain Ising-like macrospins, mimicking the atomic spins of their naturally-occurring 3-dimensional analogs \cite{realspin,ramirez,castelnovo}, but confined to a plane. They are thus realizations of the square ice vertex models solved by Wu \cite{Wu1967} and Lieb \cite{Lieb1967}, in which the exact microstate of the statistical mechanical system can be inspected using advanced microscopy methods. Until very recently, all ASIs studied have been athermal--albeit showing an effective thermodynamics \cite{2007Nisoli,Nisoli2010,entropy,Morgan2013}--since the shape anisotropy energy barrier $E_\mathrm{A}$ that must be surmounted to flip the magnetic moment of any one island is orders of magnitude larger than the thermal energy $k_\mathrm{B}T$ that can be reached experimentally. Whilst convenient for imaging studies, these athermal, arrested systems lack ergodicity and so fail to explore phase space in the manner of a true statistical mechanical system to find thermally equilibrated configurations.

\begin{figure}[t]
  \includegraphics[width=5cm]{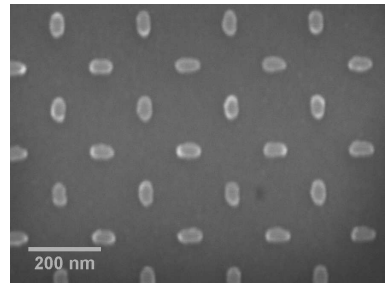}
  \caption{Scanning electron micrograph of an artificial spin ice with a lattice spacing of 240~nm, with Permalloy islands of lateral size 30~nm~$\times$~70~nm and 8~nm thickness. \label{SEM_lattice}}
\end{figure}

However, there are recent reports of thermalized ASIs, made either by heating the sample close to the Curie point of the material from which the islands are fabricated in order to drive dynamics \cite{melting}, whereupon the arrested state may be imaged upon cooling \cite{porro,zhang}, or a one-shot anneal process that occurs during fabrication has been used in the same way \cite{morgan}. Within the last year or two, studies have been carried out that have dynamically imaged real-time thermal fluctuations in artificial spin ice in the square \cite{farhansquare}, kagom\'{e} \cite{hypercube,Kapaklis2014}, and tetris ice geometries \cite{Gilbert2015}. Nevertheless the nature of the crossover from a thermally fluctuating system to the arrested,  less ergodic state has so far received little attention. Here we report on measurements of an ASI with islands of lateral size 30~nm~$\times$~70~nm \cite{Morley2015}, shown in Fig.~\ref{SEM_lattice}. By studying the time-dependence of their soft x-ray speckle scattering patterns, the ASI is shown to be thermal at room temperature and to freeze into a fully arrested state below $\sim 178$~K. The fluctuation rate follows a Vogel-Fulcher-Tammann law on cooling, implying that the frozen state is glassy in nature, which is unexpected for a system with a well-defined ground state.

\begin{figure}
  \includegraphics[width=8cm]{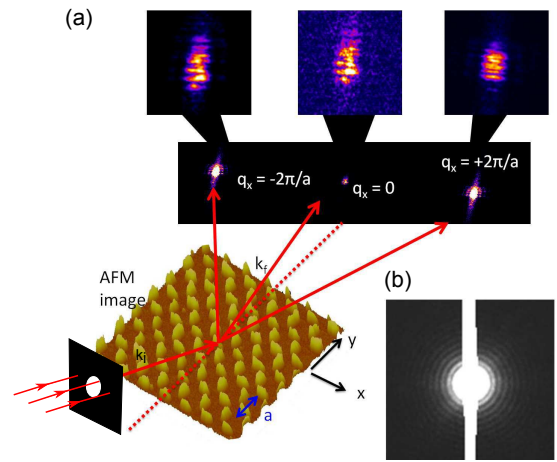}
  \caption{(Color online) Coherent soft x-ray scattering measurements. (a)~Schematic of the experimental XPCS setup, showing the incoming x-ray beam and scattered beams from the ASI array, which is represented by an atomic force microscopy image. The three main diffraction spots in the row above the specular reflection are shown here. Since the diffraction spots arise from a small, disordered region of the sample that has been coherently illuminated, they contain speckle contrast. (b)~Fraunhofer diffraction obtained using the 10~$\upmu$m pin hole and straight-through beam on to the CCD detector. \label{schem}}
\end{figure}

The ASI that we studied was fabricated using electron beam lithography with a lattice constant of 240~nm, as shown in Fig.~\ref{SEM_lattice}. A lift-off process was used, with 8~nm thick Permalloy (Ni$_{80}$Fe$_{20}$) evaporated through the resist mask to form the islands, followed by a 2~nm Al cap. The substrate was a Si wafer. The ASI array had a 2~mm~$\times$~2~mm area. Soft x-ray photon correlation spectroscopy (XPCS) coherent scattering measurements were carried out at the I10 beamline at Diamond in order to measure the time-time correlations of the magnetic fluctuations in the ASI. The experiments were carried out using circularly polarized light at the Fe $L_{3}$ (707 eV) resonance. The scattered intensity from the ASI was recorded in the reflection geometry (illustrated in Fig.~\ref{schem}(a)) using a charge coupled device (CCD) camera, mounted at a fixed scattering angle $2\theta = 9.6^{\circ}$, 80~cm from the sample, which was kept at a temperature of 223~K in order to reduce dark noise. Each image arose from a 40~ms exposure and the images were separated by the 4~s readout time of the camera. The transverse coherence length of the beam was calculated to be 14.9~$\upmu$m, and so a 10~$\upmu$m diameter pinhole was mounted 23~cm in front of the sample. Airy rings were observed when the direct beam was imaged through the pinhole, shown in Fig.~\ref{schem}(b), confirming the coherence of the beam. The samples were mounted for measurement on a temperature-controlled stage in the absence of any applied magnetic field.

Since our ASI array has a square unit cell, it produces a square pattern of diffraction spots. In Fig.~\ref{schem}, the centre of the diffracted speckle pattern corresponds to $Q_{x} = 2\pi/a$, where $a$ is the ASI lattice constant. The calculated lattice spacing for the ASI measured from the position on the array detector and the geometry was $239 \pm 2$~nm, which agrees with that measured from SEM, $243 \pm 4$~nm. In previous scattering experiments on square ASIs incoherent light has been used to determine the different sublattice contributions to the hysteresis measured on different orders of diffraction \cite{inplaneBragg} or to observe half order magnetic ground state peaks from as-grown samples \cite{Perron2013}. When coherent illumination is used then the diffraction spot decomposes into speckle, arising from the disorder in the illuminated region of the sample introducing different phase shifts into the scattered waves \cite{sutton}: this speckle is visible in the detailed views of the diffraction spots shown at the top of Fig.~\ref{schem}(a). The crux of the XPCS technique is that dynamics in the magnetic configuration is reflected in changes in the details of the speckle pattern. This effect has long been used with light at visible wavelengths to study relaxation and aging phenomena in soft matter \cite{Ciplelletti2003}. XPCS experiments utilising hard x-rays have been used to study charge density waves associated with antiferromagnetic domains in Cr \cite{shpyrko}, giving an indirect measurement of magnetic dynamics. Soft x-rays at the $M_5$ resonance of Ho revealed antiferromagnetic domain fluctuations in a thin film of that metal \cite{Holmium}, whilst the jamming of spiral magnetic domains in a Y-Dy-Y trilayer was revealed in the stretched exponential correlations studied using soft X-rays at the Dy $M_5$ resonance \cite{jamming}. There have also been speckle scattering studies at the Co $L_3$ resonance using small-angle X-ray scattering geometry to study the effects of disorder on the domain pattern of multilayer perpendicular Co/Pt films in response to a field \cite{Pierce2007}.

\begin{figure}[t]
  \includegraphics[width=7.5cm]{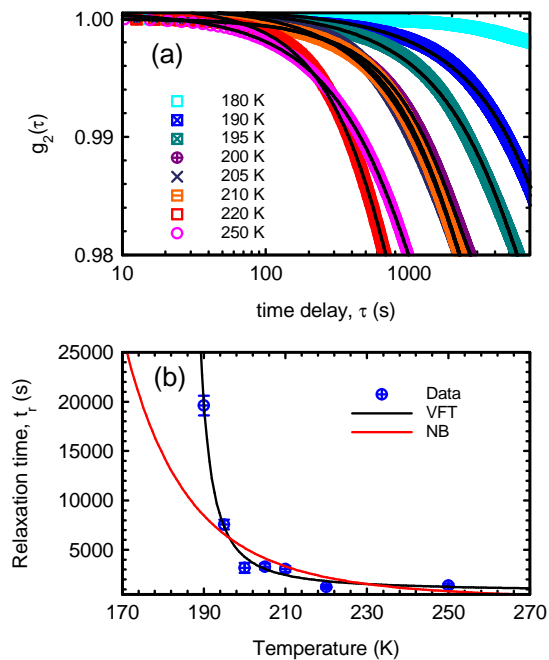}
  \caption{(Color online) XPCS results. (a)~Normalized $g_2(\tau)$ functions at various temperatures. The lines are fits to Eq. \ref{kww}. (b)~Relaxation times $t_\mathrm{r}$ as a function of temperature, fitted by a Vogel-Fulcher-Tammann (VFT) law. Also plotted is the superparamagnetic N\'{e}el-Brown (NB) law, which fits poorly.}
  \label{fig:g2temp}
\end{figure}

As expected, measurements taken at an energy of 700~eV, below the Fe $L_3$ resonance, showed no change beyond random noise fluctuations (see Supplementary Movie 1), since the ASI physical structure is static. On tuning to the $L_3$ resonance at 709~eV, magnetic sensitivity is achieved and the speckle reconfigures as time passes, since the magnetic state of the sample is reconfiguring under thermal activity (see Supplementary Movie 2). The XPCS measurements were carried out at different temperatures to drive the thermal fluctuations at different rates. In order to quantify the time-dependent behaviour, we calculated the intensity-intensity temporal autocorrelation function \cite{Ciplelletti2003},
\begin{equation}
  g_{2}(\mathbf{Q},\tau) = \frac{\langle I(\mathbf{Q},t^\prime) I(\mathbf{Q},t^\prime+ \tau) \rangle_{t^\prime}}{\langle I(\mathbf{Q},t^\prime)^2 \rangle_{t^\prime}} = 1 + A \left| F(\mathbf{Q},\tau) \right|^2,
\end{equation}
where $I(\mathbf{Q},t^\prime)$ is the intensity at wave vector $\mathbf{Q}$ at a time $t^\prime$, $\tau$ is the time delay, and $\langle ... \rangle_{t^\prime}$ indicates a time average. $F(\mathbf{Q},t^\prime)$ is the so-called intermediate scattering function, and $A$ is a measure of the degree of speckle contrast. The $g_2$ function was calculated for each pixel within the speckle pattern of the Bragg peak, and averaged over all such pixels to give the autocorrelation $g_{2}(\tau)$ at each temperature, shown normalized to an initial value in Fig.~\ref{fig:g2temp}(a).

These $g_{2}(\tau)$ curves were fitted with a heterodyne model, since the experimental setup means that the fluctuating magnetic signal is mixed with a static signal that comes from the structural scattering of the array \cite{Sinha2014}. This forms the equivalent of a reference beam in the measurement, leading to the following modified form of the Kohlrausch-Williams-Watts \cite{Kohlrausch1854,Williams1970} intermediate scattering function \cite{livet}:
\begin{equation}
  g_{2}(\tau) = 1 + A \cos (\omega\tau) \exp\left( -(\tau/t_\mathrm{r})^{\beta} \right) \label{kww}
\end{equation}
where $t_\mathrm{r}$ is the characteristic relaxation time, $\beta$ is a stretching exponent, and $\omega$ is an oscillation frequency associated with the heterodyne mixing. The value of $\omega$ was found to be in the range $\approx 0.001$-0.002~rads$^{-1}$, corresponding to a time period, 2$\pi$/$\omega$ = 3,140~-~15,700~s ($\approx 1$-4~hrs), which correlates well with the total measurement time. The speckle contrast coefficient, $A$, was found to be in the range 0.02-0.05, which agrees with that found from the contrast of the Airy pattern \cite{beutier}. The fitted value of $\beta$ in all our measurements was $1.0 \pm 0.1$, indicating equilibrated behavior \cite{shpyrko}. The function is rather flat at 180~K, and therefore the relaxation time must be much longer than the time of measurement ( $\approx 3$~hrs) at that temperature, and so cannot be accurately determined. The correlation function drops off more and more quickly with increasing temperature, as can be seen from the initial slope and the drop-off in Fig.~\ref{fig:g2temp}(a). At room temperature the system fluctuates too quickly and the speckle appears blurred due to the insufficient time resolution of our CCD acquisition.

For temperatures between 190 and 250~K, $t_\mathrm{r}$ lies in the experimentally accessible range between these two extremes, shown in Fig.~\ref{fig:g2temp}b. The simplest thermal activation behavior in a collection of magnetic nanoparticles is the Arrhenius-type N\'{e}el-Brown (NB) law expected for a non-interacting superparamagnetic system, $t_\mathrm{r} = \tau_{0} \exp (T_\mathrm{A}/T)$, where $T_\mathrm{A} = E_\mathrm{A}/k_\mathrm{B}$ is an activation temperature and $\tau_{0}$ is an activation time. As can be seen in the figure, this cannot be fitted to the data at all well. On the other hand, a Vogel-Fulcher-Tammann (VFT) law \cite{Vogel1921,Fulcher1925,Tammann1926} captures the low temperature detail much more accurately: the data were fitted with the expression
\begin{equation}
  t_\mathrm{r} = \tau_{0} \exp \left( \frac{T_\mathrm{A}}{T-T_{0}} \right), \label{vf}
\end{equation}
where $T_{0}$ is the freezing temperature. The fit yields $T_\mathrm{A} =  40 \pm 10$~K and $T_\mathrm{0} =  178 \pm 5$~K.

The defining feature of a VFT law is the freezing temperature $T_0$, below which interactions in the system prevent any relaxation or fluctuation. Whilst the VFT law has been found to fit many different types of data sets very well, such as spin glasses, super-cooled organic liquids, metallic liquids, and glassy (bio)polymer systems \cite{Angell1995,cyrot}, it is nevertheless an empirical law not based on any underlying microscopic picture. Models that attempt to provide a VFT-like temperature dependence include--among others--ones based on a time-dependent percolation process \cite{cyrot}, or on the energy distribution of the depth of coupled traps \cite{Bouchaud1995}. In the spin glass theory of Shtrikman and Wohlfarth (S-W) \cite{shtrik}, the random exchange interactions between the positionally disordered spins are represented by a mean magnetic interaction field $H_\mathrm{i}$. The magnetostatic interactions in our ASI are real magnetic fields, making the S-W theory a natural one to adapt to our system. In it, the freezing temperature $T_0$  is determined by the characteristic interaction energy $E_\mathrm{i}$. Using the results from the VFT fit, we find $E_{\mathrm{i}} = \sqrt{k_{\mathrm{B}}^{2} T_{0} T_{\mathrm{A}}} \approx 1.2\times 10^{-21}$~J.

We can estimate the scale of magnetostatic interactions in our ASI using a point dipole model to determine the S-W interaction energy:
\begin{equation}
  E_\mathrm{i} = \mu_0 H_\mathrm{i} m_\mathrm{island} = \frac{\mu_{0} m_{\mathrm{island}}^{2}}{2 \pi a^{3}},
\end{equation}
where $m_\mathrm{island} = M_\mathrm{s}V$ is the magnetic moment of an island of volume $V$ and the lattice constant $a = 240$~nm. Assuming the nominal island size and the bulk magnetization of Permalloy ($M_\mathrm{s} = 860$~kA/m) yields $E_\mathrm{i} = 3.0\times 10^{-21}$~J, which is of the same order of magnitude as the experimental value but overestimated by about a factor of three. Part of the overestimate may be due to the extrinsic effect of the reduction in effective $M_\mathrm{S}$ \cite{Kapaklis2014} or $V$ \cite{Le2015} due to oxidation. An intrinsic contribution to the overestimate is the fact that the total interaction energy arises from the sum of several frustrated interactions from all the island's neighbors, leading to partial cancellation of the energy value given above. Thus, the magnetostatic interactions in the system are clearly of the right scale to break the Arrhenius-like behaviour and give rise to a VFT-like freezing of the fluctuations as $T_0$ is approached.

Now we turn to a discussion of the activation temperature $T_\mathrm{A}$. We can estimate the energy barrier $\Delta E$ for Stoner-Wohlfarth coherent rotation of a single island, again assuming the bulk magnetization of Permalloy, using $\Delta E = KV = \ln (f_0 t_{\mathrm{m}})k_{\mathrm{B}}T_{\mathrm{A}}$, with the shape anisotropy constant $K = \frac{1}{2}\mu_0 \Delta{\cal D} M^2$, and the difference in demagnetizing factor along the two relevant directions $\Delta{\cal D} \approx 0.1$ based on the island geometry \cite{osborn}. Here the measurement time, $t_{\mathrm{m}}$, is taken as two hours, a typical value for $2 \pi / \omega$, and we assume $f_0 = 10$~GHz. This yields $T_{A} \approx 1700$~K. This is far larger than the fitted $T_{\mathrm{A}}$ of 40~K, ruling out this simple picture.

\begin{figure}[t]
  \includegraphics[width=7.5cm]{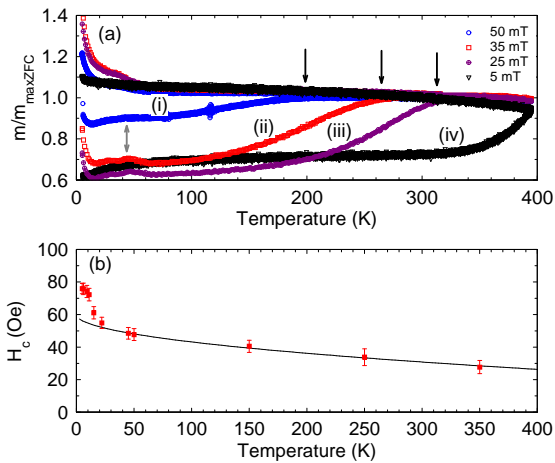}
  \caption{(Color online) (a)~ZFC and FC magnetization measurements under different applied magnetic probe fields of (i) 50~mT, (ii)~35~mT, (iii)~25~mT and (iv)~5~mT. The magnetic moment, $m$, has been normalised to the peak in the ZFC curve for each field ($m_{\mathrm{maxZFC}}$) for comparison. The average blocking temperature, $T_{\mathrm{B}}$, is also defined by this point and is indicated by the black arrows ($T_{\mathrm{B}} > 400$~K at 5~mT). The grey arrow indicates the $\sim 40$~K feature in the ZFC data. The field was applied along $[01]$ direction of the ASI lattice. (b)~The temperature dependence of the coercivity measured along $[01]$. The line is a fit to Sharrock's equation, the departure from which is clear below 40~K.
  \label{fig:zfcfc}}
\end{figure}

In order to seek other evidence of thermally-activated processes that might cast light on this large discrepancy, standard zero-field cooled (ZFC) and field-cooled (FC) protocols were carried out on the samples using SQUID-vibrating sample magnetometry. First, the sample was heated to 395~K and a large negative saturating field applied and removed. The sample was then cooled to 5~K without any applied field. Next, a probe field was applied and the magnetic moment was measured during heating, which is the lower curve in each measurement in Fig.~\ref{fig:zfcfc}(a). Last, the sample was measured again during cooling back down with the probe field still applied. The peak in the ZFC, close to where the curves bifurcate, is defined as the blocking temperature, $T_{\mathrm{B}}$, and marked with a black arrow. For example, the data with probe field $\mu_{0}H = 50$~mT has $T_{\mathrm{B}}(50~\mathrm{mT}) = 200 \pm 10$~K. The blocking temperature systematically rises for smaller probe fields, and exceeds our experiemntal limit of 400~K at 5~mT. This behavior can be extrapolated to the case of zero applied probe field, using $T_{\mathrm{B}} \propto H^{2/3}$ \cite{ElHilo1992}, which gives $T_{\mathrm{B}}(0~\mathrm{mT}) = 460 \pm 20$~K. This is about a factor of four lower than that expected from the sample parameters and nominal volume. Whilst reduced magnetization or volume can explain some of this discrepancy, it seems clear once again that a pure coherent rotation mechanism is unlikely to be strictly followed.

Nevertheless, there is another feature in the magnetometry data: there is an initial increase in moment up to 40~K, marked with a grey double-headed arrow, visible in all the ZFC curves. This signifies a much lower energy scale for some magnetic relaxation process, one that is strikingly similar to the $T_\mathrm{A}$ found from the VFT fit. As previously shown by Ozatay et al. \cite{Ozatay2008}, the properties of Py can change below 40~K when native oxidation has occurred. They showed in elements of a similar size that such oxidation is likely around the unprotected edges and can have significant effects on the reversal properties, where they observed an upturn in coercivity below this temperature.

We performed the same measurement and have plotted the coercivity of our sample as a function of temperature in Fig.~\ref{fig:zfcfc}(b). The same characteristic features as those seen in Ref.~\onlinecite{Ozatay2008} are present. The line in the figure is a fit to Sharrock's equation \cite{sharrock}, which describes the temperature dependence expected for the dynamic coercivity of magnetic particles. Ozatay et al. showed that for conformally capped islands this feature disappears and the coercivity can be fit to Sharrock's equation for all temperatures, whereas here and for their Py particles with a native oxide around the edges there is a sharp departure from this at around 40~K. They attribute this to the modification of the local anisotropy which arises from the exchange coupling between the antiferromagnetic oxide edges and the ferromagnetic Py. This results in different local anisotropy axes which create pinning points for the magnetization, and allow non-uniform reversal modes to become energetically accessible. This allows the system to sample a distribution of metastable states only once a temperature of $T_\mathrm{A} = 40$~K is exceeded.



To summarize, on cooling an artificial spin ice we have observed a dramatic lengthening of the relaxation time as measured by magnetic XPCS. The system slows abruptly as it crosses over from thermal equilibration to an athermal, frozen state. This crossover can be described by a Vogel-Fulcher-Tammann law, which is typically used to describe glassy systems. The VFT freezing temperature $T_0 \approx 178$~K can be accounted for by the magnetostatic interaction strength through Shtrikman-Wohlfarth theory. The activation temperature $T_\mathrm{A} \approx 40$~K arising from a fit of this law implies a much lower energy barrier to reversal than is expected from a single-domain coherent rotation picture. The value of 40~K appears to be determined by the onset of fluctuating magnetic states within the islands coupled to magnetic oxides at the edges of the islands. 

The glass-like freezing is remarkable since the square ice system possesses a well-defined ground state, unlike a conventional spin glass. This raises the question of the true nature of the glassy state that our artificial spin ice freezes into. Related VFT-like freezing behaviour has recently been observed in the pyrochlore spin ice Dy$_2$Ti$_2$O$_7$ \cite{Kassner2015}, prompting speculation about this representing many-body localization of spins in a translationally-invariant quantum system \cite{DeRoeck2014,Yao2014,Schiulaz2015}. This is unlikely to be the case here since the artificial systems of the type we have studied consist of classical macrospins, ruling out explicitly quantum behavior, implying that some other unusual glass state is obtained in artificial spin ices once ergodicity is broken.

\begin{acknowledgments}
This work was supported by the EPSRC (grant numbers EP/J021482/1, EP/I000933/1, and EP/L002922/1). Research carried out in part at the Center for Functional Nanomaterials, Brookhaven National Laboratory, which is supported by the U.S. Department of Energy, Office of Basic Energy Sciences, under Contract No. DE-AC02-98CH10886. We would like to thank Diamond Light Source for beamtine and S. S. Dhesi for the loan of the CCD camera.
\end{acknowledgments}

\bibliography{MorleyThesis}

\end{document}